\documentclass{aastex631}

\usepackage{amsmath}
\usepackage{url}
\usepackage{bm}
\usepackage{xcolor}
\usepackage{subfigure}
\usepackage{graphicx}

\begin{document}

\title{New $^{26}$P($p$,$\gamma$)$^{27}$S thermonuclear reaction rate and its astrophysical implication in rp-process}

\author{S.Q.~Hou}
\affiliation{CAS Key Laboratory of High Precision Nuclear Spectroscopy, Institute of Modern Physics, Chinese Academy of Sciences, Lanzhou 730000, China}
\affiliation{School of Nuclear Science and Technology, University of Chinese Academy of Sciences, Beijing 100049, China}
\affiliation{NuGrid Collaboration, \url{http://www.nugridstars.org}}

\author{J.B.~Liu}
\affiliation{CAS Key Laboratory of High Precision Nuclear Spectroscopy, Institute of Modern Physics, Chinese Academy of Sciences, Lanzhou 730000, China}
\affiliation{School of Nuclear Science and Technology, University of Chinese Academy of Sciences, Beijing 100049, China}

\author{T. C. L. Trueman}
\affiliation{Konkoly Observatory, Research Centre for Astronomy and Earth Sciences, Hungarian Academy of Sciences, H-1121 Budapest, Hungary}
\affiliation{E. A. Milne Centre for Astrophysics, University of Hull, Kingston upon Hull HU6 7RX, United Kingdom}
\affiliation{NuGrid Collaboration, \url{http://www.nugridstars.org}}

\author{J.G.~Li}
\affiliation{CAS Key Laboratory of High Precision Nuclear Spectroscopy, Institute of Modern Physics, Chinese Academy of Sciences, Lanzhou 730000, China}
\affiliation{School of Nuclear Science and Technology, University of Chinese Academy of Sciences, Beijing 100049, China}

\author{M.~Pignatari}
\affiliation{Konkoly Observatory, Research Centre for Astronomy and Earth Sciences, Hungarian Academy of Sciences, H-1121 Budapest, Hungary}
\affiliation{E. A. Milne Centre for Astrophysics, University of Hull, Kingston upon Hull HU6 7RX, United Kingdom}
\affiliation{Joint Institute for Nuclear Astrophysics, Center for the Evolution of the Elements, Michigan State University, East Lansing, Michigan 48824, USA}
\affiliation{NuGrid Collaboration, \url{http://www.nugridstars.org}}

\author{C.~Bertulani}
\affiliation{Department of Physics and Astromomy, Texas A\&M University-Commerce, Commerce, Texas 75429, USA }

\author{X.X.~Xu}
\affiliation{CAS Key Laboratory of High Precision Nuclear Spectroscopy, Institute of Modern Physics, Chinese Academy of Sciences, Lanzhou 730000, China}
\affiliation{School of Nuclear Science and Technology, University of Chinese Academy of Sciences, Beijing 100049, China}

\correspondingauthor{S.Q.~Hou}
\email{sqhou@impcas.ac.cn}



\begin{abstract}
Accurate nuclear reaction rates for $^{26}$P($p$,$\gamma$)$^{27}$S are pivotal for a comprehensive understanding of rp-process nucleosynthesis path in the region of proton-rich sulfur and phosphorus isotopes. 
However, large uncertainties still exist in the current rate of $^{26}$P($p$,$\gamma$)$^{27}$S because of the lack of the nuclear mass and the energy level structure information of $^{27}$S. We reevaluate this reaction rate using the experimentally constrained $^{27}$S mass, together with the shell-model predicted level structure. It is found that the $^{26}$P($p$,$\gamma$)$^{27}$S reaction rate is dominated by a direct-capture (DC) reaction mechanism despite the presence of three resonances at E = 1.104, 1.597, 1.777 MeV above the proton threshold in $^{27}$S. The new rate is overall smaller than the other previous rates from Hauser-Feshbach statistical model by at least one order of magnitude in the temperature range of X-ray burst interest. In addition, we consistently update the  photodisintegration rate using the new $^{27}$S mass. The influence of new rates of forward and reverse reaction in the abundances of isotopes produced in rp-process is explored by post-processing nucleosynthesis calculations. The final abundance ratio of $^{27}$S/$^{26}$P obtained using the new rates is only 10\% of that from the old rate. The abundance flow calculations show the reaction path $^{26}$P($p$,$\gamma$)$^{27}$S($\beta^+$,$\nu$)$^{27}$P is not as important as thought previously for producing $^{27}$P . The adoption of the new reaction rates for $^{26}$P($p$,$\gamma$)$^{27}$S only reduces the final production of aluminum by 7.1\%, and has no discernible impact on the yield of other elements.
\end{abstract}



\section{Introduction} \label{sec:intro}

The rapid proton capture process, hereafter rp-process, typically occurs when hydrogen fuel is ignited under highly degenerate conditions in explosive events on the surface of compact objects like white dwarfs (novae) \citep{Truran82,Truran90,Shankar90,Starrfield93} and neutron stars (X-ray bursts) \citep{Wallace81,Taam85,Taam93,Lewin93}. During this process, the consecutive proton captures onto stable nuclei can produce nuclei far away from the stability line, even approaching the proton drip line \citep{Jordi10}. Naturally, this process competes with the $\beta^+$-decays and reverse photodisintegration reactions. For a given nucleus, if proton capture on it is inhibited by the strong reverse reaction rate and subsequently has to wait for a slower $\beta^+$-decay, the nucleus is often termed a waiting point nucleus \citep{Schatz98}.

In the nuclide chart, the nucleus $^{27}$S is located on the proton-rich side, far beyond the valley of stability. During the rp-process, $^{27}$S is synthesized via two successive  proton captures on the waiting-point nuclei $^{25}$Si which is characterized by a ($p$,$\gamma$)-($\gamma$,$p$) equilibrium between $^{25}$Si and $^{26}$P. However, regarding the identification of the waiting-point nuclei $^{25}$Si, it is actually affected by the net reaction flow leaking out of the equilibrium through $^{26}$P($p$,$\gamma$)$^{27}$S. From the principle of detailed balance, we know that the forward and reverse rates for an arbitrary reaction can be mutually converted via a Q-value-dependent exponential term. Therefore,  accurate $^{26}$P($p$,$\gamma$)$^{27}$S rates and the nuclear masses of $^{26}$P and $^{27}$S are of great importance for determining the degree to which $^{25}$Si is a waiting-point nucleus and thus a better understanding of rp-process reaction path in the region around $^{26}$P.

In \cite{Parikh13}, by using nucleosynthesis calculations it is shown that the nuclear mass of $^{27}$S is essential to quantify the abundance flows proceeding the waiting-point nuclei $^{25}$Si. However, it also pointed out that the nuclear mass uncertainty of $^{27}$S from AME2003\citep{Audi03} used in their calculation cannot reach the expected accuracy. Regarding the issue of how much the actual leakage from $^{26}$P occurs via $^{26}$P($p$,$\gamma$)$^{27}$S, it is determined by the degree of competition between the forward reaction $^{26}$P($p$,$\gamma$)$^{27}$S and the reverse process $^{27}$S($\gamma$,$p$)$^{26}$P under the astrophysical condition of X-ray burst. Therefore, the accurate treatment of forward and reverse reaction rates for $^{26}$P($p$,$\gamma$)$^{27}$S is crucial for a better understanding of the rp-process nucleosynthesis path in the region of nuclei with mass number A = 26-27.

Concerning the reaction $^{26}$P($p$,$\gamma$)$^{27}$S, all of the three currently existing rates are from the theoretical calculation by Hauser-Feshbach statistical model \citep{Rauscher00,Cyburt10}. However, the statistical model is only suitable for the case of the high energy level density in compound nuclei near the proton threshold \citep{Rauscher00}. Therefore, both the shell model prediction of $^{27}$S and the measured mirror states in $^{27}$Na indicate the statistical model is not a good choice for the prediction of the $^{26}$P($p$,$\gamma$)$^{27}$S rate. The respective photodisintegration rate can be derived from the forward rate if the reaction Q-value is available \citep{Herndl95,Iliadis15}.

The nuclear masses of the involved nuclei $^{26}$P and $^{27}$S used in the investigation by Parikh are taken from AME2003 \citep{Audi03}, where the corresponding masses are evaluated as 10970 keV and 17540 keV, both with an uncertainty of 200 keV. Differing from the case of the $^{26}$P mass remaining constant as in AME2003, the nuclear mass of $^{27}$S varies from 17030(400) keV to 17490(400) keV in several different recent versions of atomic mass evaluations \citep{Wang12,Wang16,Wang21}. Because all of these are theoretically deduced from trends in the mass surface, it is difficult to tell which one can represent the actual mass of $^{27}$S. Thanks to the recent $\beta$-decay spectroscopy experiment of $^{27}$S, a more precise $^{27}$S mass excess of 17678(77) keV was proposed based on the measured $\beta$-delayed two-proton energy and the Coulomb displacement energy relations \citep{Sun20}. The new mass not only affects sensitively the reverse reaction rate of $^{26}$P($p$,$\gamma$)$^{27}$S via an exponential term of exp(-Q/kT) which exists between the forward and reverse reaction rates, but it also has an important influence in the forward reaction rate since the new $^{27}$S mass will lead to a huge alteration of the resonance contribution. In this work, we reevaluate the $^{26}$P($p$,$\gamma$)$^{27}$S reaction rate using the new mass of the involved nuclei, in a combination of shell model predictions of the energy structure of $^{27}$S. Meanwhile, we also explore the impact of the new rates on the nucleosynthesis path in the rp-process, and the role of nuclear physics uncertainties in abundance prediction. 
    
The paper is structured as follows. Section 2 introduces the shell-model calculation on the energy level structure of $^{27}$S and the astrophysical reaction rate calculation. We investigate the effect of new rates for $^{26}$P($p$,$\gamma$)$^{27}$S to the rp-process nucleosynthesis in Section 3. The conclusions are discussed in Section 4.

\section{$^{26}$P($p$,$\gamma$)$^{27}$S Reaction rate} \label{sec:reaction rate}
The energy structure of compound nucleus $^{27}$S remains unclear currently. In light of its mirror nucleus $^{27}$Na with a low energy level density, it is worth investigating the reaction rate of $^{26}$P($p$,$\gamma$)$^{27}$S, especially the contributions from the resonances near threshold. Below we will make a detailed study on this reaction based on the shell model and resonant reaction theory.

\subsection{Shell-model calculation}
\begin{figure}[ht!]
\begin{center}
\includegraphics[width=0.7\textwidth]{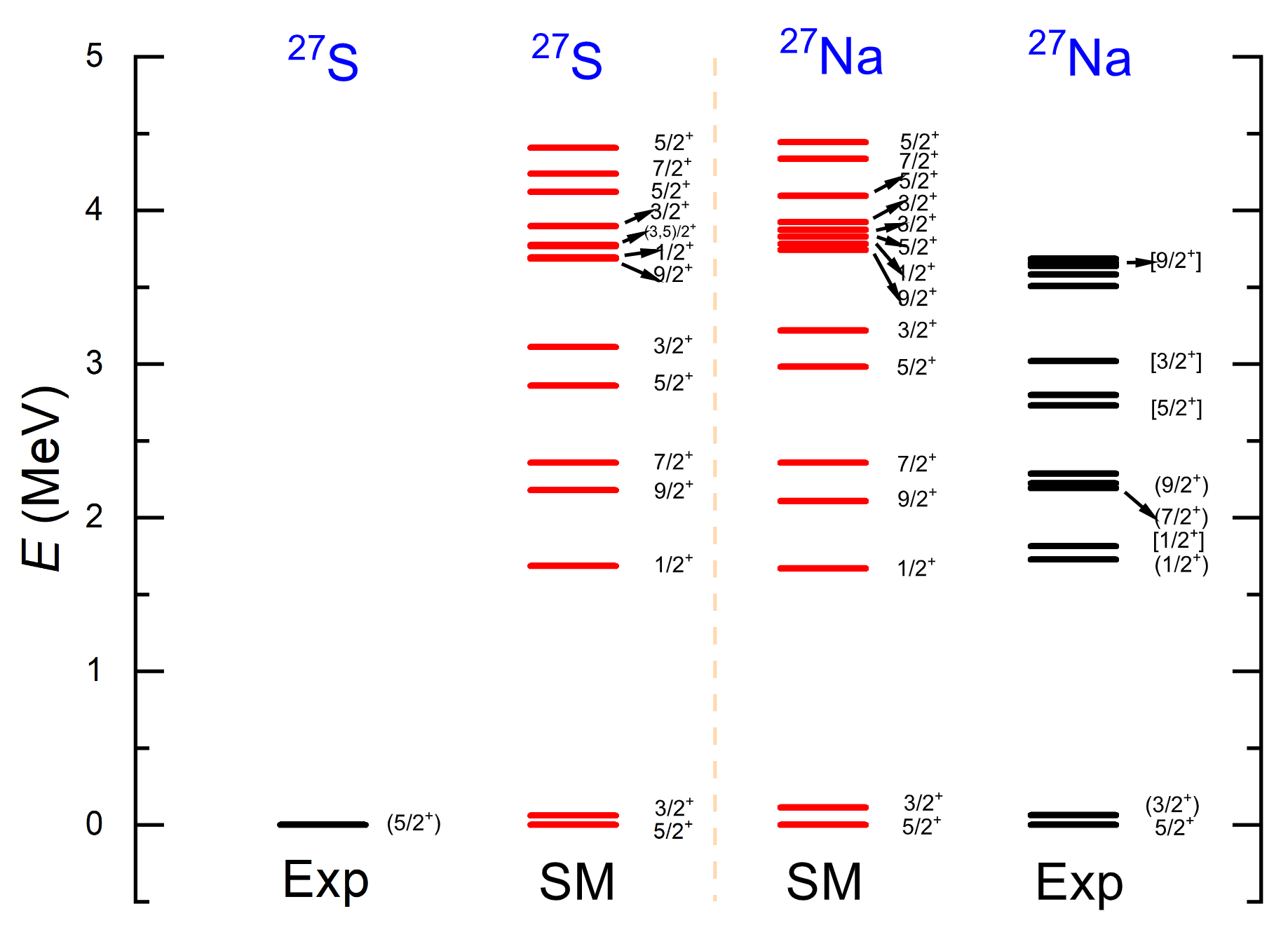}
\end{center}
\caption{\label{fig1} (Color online) Comparison of the experimental and theoretical excitation energies for the mirror nuclei $^{27}$S and $^{27}$Na, SM is the result of Shell-model, Exp is the result from the experiment.
\label{level scheme}}
\end{figure}

Regarding the $^{27}$S energy structure, the only existing information is on the ground state \citep{Shams11}. Despite this, the energy levels of its mirror nucleus $^{27}$Na have been obtained from the experiment of the $\beta^{-}$ decay of $^{27}$Ne by \cite{Tri06} -- the spin and $\gamma$ transition probabilities of these excited state of interest were not given. Therefore, we have to choose a shell-model to obtain the necessary energy structure information used to calculate the resonant reaction rates. We calculate the $^{27}$S energy levels using the KSHELL shell-model code \citep{Shimizu19}, and the $sd$-shell model space involving the  $\pi0d_{5/2}$, $\pi0d_{3/2}$, $\pi1s_{1/2}$, $\nu0d_{5/2}$, $\nu0d_{3/2}$ and $\nu1s_{1/2}$ valence orbits. Here, $\pi$ denotes proton, $\nu$ denotes neutron. 
The new isospin-breaking USDC interaction was used in the present works \citep{Brown20}.
The calculated results on the energy levels of $^{27}$S and the mirror nucleus $^{27}$Na are shown in Fig.\ref{level scheme}. For comparison, the available experimental energy levels of this pair of mirror nuclei are added as well. We can see that the predicted low energy levels of $^{27}$Na by shell-model basically match the experimental result. In addition, the critical information used to calculate resonant reaction rates, such as reduced transition probabilities $B(M1)$ and $B(E2)$ and the spectroscopic factor of $^{26}$P($p$,$\gamma$)$^{27}$S, are also obtained through shell-model.

\subsection{Reaction-rate calculation}
The thermonuclear $^{26}$P($p$,$\gamma$)$^{27}$S reaction rate is the incoherent sum of all resonant and nonresonant capture contributions. It is well known that only the resonances located in the energy window of astrophysical interests (called Gamow window) contribute significantly to the reaction rate.  For a narrow isolated resonance, the resonant reaction rate can be calculated using the below expression \citep{Herndl95, He17}
\begin{equation}
N_A\left\langle\sigma\nu\right\rangle_r =1.5394\times10^{11}\times(\mu T_9)^{-3/2}\times\omega\gamma
\times{\rm exp}\left( -\frac{11.605E_r}{T_9}\right)({\rm cm^3s^{-1}mol^{-1}})\text{,}
\end{equation}
where $N_A$ is Avogadro’s constant, the reduced mass $\mu$ is given by $A_TA_p/(A_p+A_T)$, and $A_p=1$ and $A_T=26$ are the mass numbers of the proton and $^{26}$P, respectively. $T_9$ is the temperature in units of gigakelvin (GK), and both the resonance energy $E_r$ and resonance strength $\omega\gamma$  are given in units of MeV. Here, the resonance strength is defined as
\begin{equation}
\omega\gamma=\frac{2J_r+1}{(2J_p+1)(2J_T+_1)}\frac{\Gamma_p\times\Gamma_\gamma}{\Gamma_{tot}}\text{,}
\end{equation}
where $J_r$ is the spin of the resonance, $J_p=1/2$ is the spin of a proton, and $J_T$ is the spin of $^{26}$P. $\Gamma_p$ and $\Gamma_\gamma$ are the proton-decay width and $\gamma$-decay width, respectively. The total width $\Gamma_{tot}$ of the resonance is thought to be the sum of its proton width ($\Gamma_p$) and $\gamma$-decay width ($\Gamma_\gamma$), $\Gamma_{tot}=\Gamma_p+\Gamma_\gamma$. The proton width can be expressed as \citep{Herndl95}
\begin{equation}
\Gamma_p=C^2S\times\Gamma_{sp}\text{,}
\end{equation}
with $C^2S$ denoting the corresponding spectroscopic factor for a particular state, and $\Gamma_{sp}$ being the single-particle width, which can be obtained by using
\begin{equation}
\Gamma_{sp}=\frac{3\hbar^2}{\mu R^2}P_l(E)\text{,}
\end{equation}
 where $R=r_0\times(1+A_T^{\frac{1}{3}})$ fm is the nuclear channel radius, and $P_l$ is the penetrability factor
\begin{equation}
P_l(E)=\frac{kR}{F_l^2(E)+G_l^2(E)}\text{,}
\end{equation}
Here, $k$ is the wave number, $R$ the channel radius, and $F_l$ and $G_l$  the standard Coulomb functions \citep{Hou15}.

The $\gamma$-decay widths are obtained from electromagnetic reduced transition probabilities $B(\Omega L;J_i\rightarrow J_f)$ ($\Omega$ stands for electric or magnetic), which carry the nuclear structure information of the resonance states and the final bound states. The reduced transition probabilities were computed within the framework of the shell model. The corresponding $\gamma$-decay widths for the most contributed transitions (\textit{M}1 and \textit{E}2) can be expressed as \citep{Herndl95}
\begin{equation}
\begin{aligned}  
\Gamma_{M1}[eV]&=1.16\times10^{-2}E_\gamma^3[MeV]B(M1)[\mu^2_N]\text{, and} \\
\Gamma_{E2}[eV]&=8.13\times10^{-7}E_\gamma^5[MeV]B(E2)[e^2fm^4]\text{,}
\end{aligned}
\end{equation}

\begin{figure}[ht!]
\begin{center}
\includegraphics[width=0.7\textwidth]{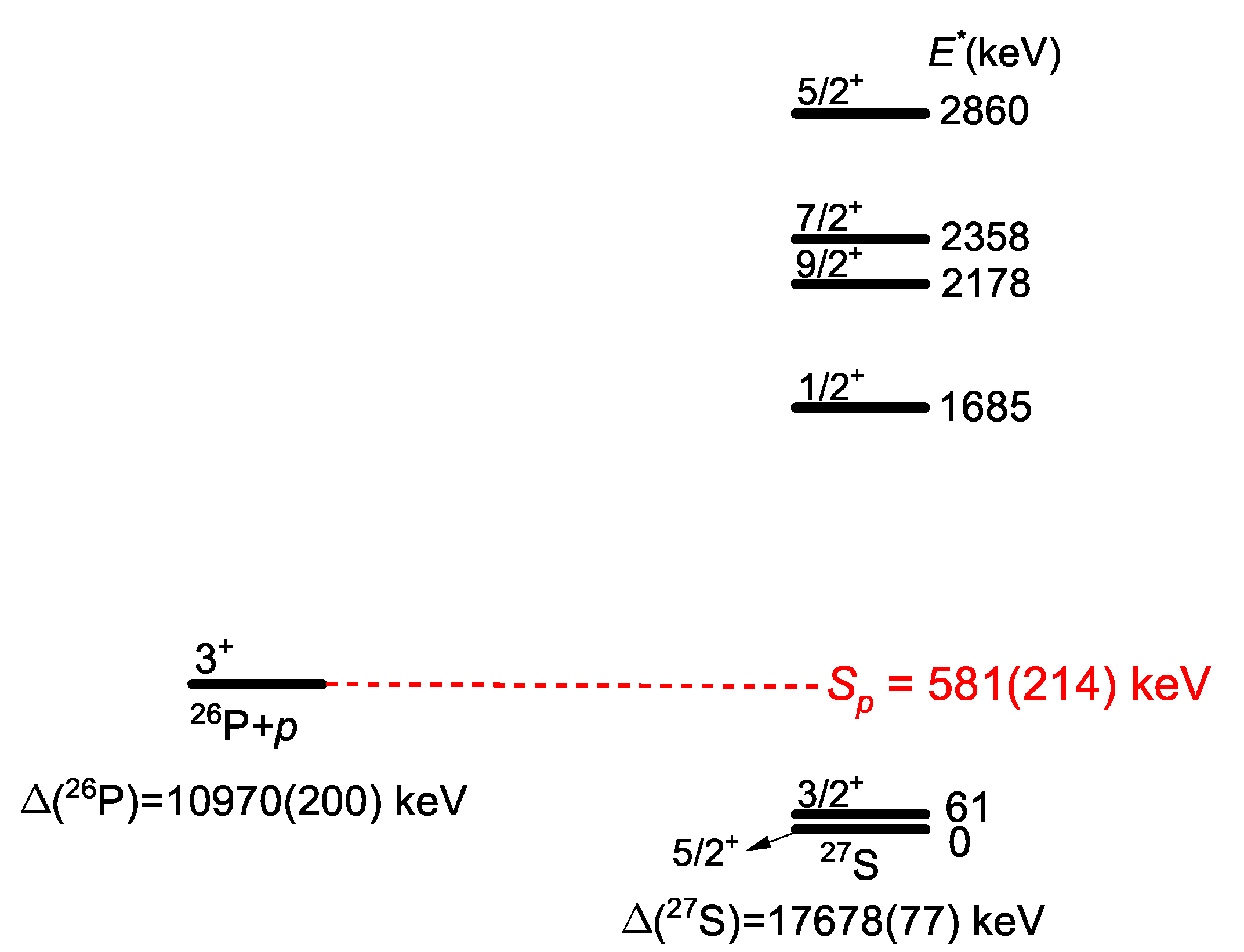}
\end{center}
\caption{(Color online) Simplified level scheme of $^{27}$S. The drawing is not to scale. The mass excesses and single proton separation energy($S_p$) are from AME2020 \citep{Wang21} and \cite{Sun19}, whereas the energies are from the shell-model calculation.
\label{Sp_xu}}
\end{figure}
\begin{figure*}[!htb]
\begin{center}
\includegraphics[width=0.6\textwidth]{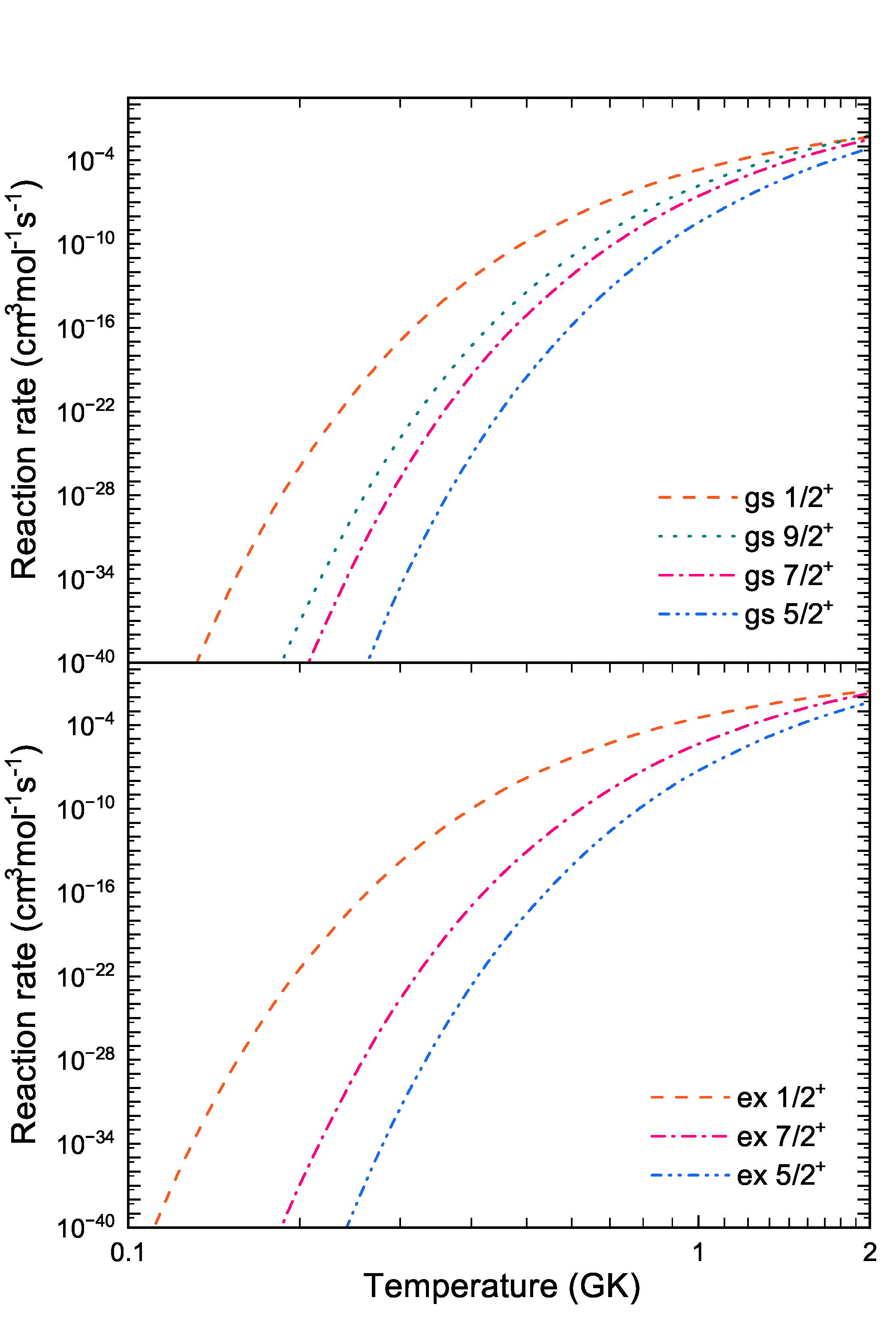}
\end{center}
\caption{(Color online) The contributions of various individual resonances to the $^{26}$P($p$,$\gamma$)$^{27}$S reaction rate as functions of temperature. In the legend, resonances are labeled with their spin and parity in $^{27}$S. The upper panel shows contributions from ground-state capture; the lower panel shows contributions from capture on the first excited state in $^{26}$P.}
\label{res_four}
\end{figure*}

In a stellar plasma, the low-lying excited states of the target nucleus are thermally populated, and might have considerable influence on the real astrophysical reaction rate. Thus, the rate contribution from the first excited state of 164 keV in $^{26}$P is also taken into our consideration because of its low excitation energy. In this work, the mass excesses of $^{26}$P and $^{27}$S are taken from AME2020 \citep{Wang21} and \cite{Sun19}, where the specific values are 10970(200) keV and 17678(77) keV, respectively. Then the proton separation energy ($S_p$) is fixed to be 581(214) keV. The four resonances closest to the proton threshold are schematically plotted in Fig.\ref{Sp_xu}. We calculate the individual reaction rate contribution for the above four resonances and plot them separately in Fig.\ref{res_four}. Since the contribution from resonance at 2.86 MeV(5/2$^+$) is negligible, we here consider three states of $^{27}$S at the resonance energies $E_r$= 1.104 MeV(1/2$^+$), 1.597 MeV(9/2$^+$), 1.777 MeV(7/2$^+$). All the relevant information for the three states is summarized in Table \ref{2} -- the upper part of the table is for ground-state capture and the lower part is for capture on the first excited state in $^{26}$P.  Columns 5-7 refer to the spectroscopic factors of the corresponding $0d_{5/2}$, $1s_{1/2}$, and $0d_{3/2}$ orbits.

\begin{table}
\scriptsize
\caption{Parameters for the present $^{26}$P($p$,$\gamma$)$^{27}$S resonant rate calculation. Listed are  excitation energy $E_x$, center-of-mass resonance energy $E_r$, 
 spin and parity $J^{\pi}$, spectroscopic factors $C^2S$,  $\gamma$-decay width $\Gamma_\gamma$, proton-decay width $\Gamma_p$, and the resonance strength $\omega\gamma$. The upper part is for ground-state capture; the lower part is for capture on the first excited state in $^{26}$P.} \label{2}
\begin{tabular}{lccccccccc}
\hline \hline
$E_x$(MeV) & $E_r$(MeV) &$J^{\pi}$ &$l$ & $C^2S_{5/2}$ &$C^2S_{1/2}$ & $C^2S_{3/2}$ & $\Gamma_\gamma$(eV) & $\Gamma_p$(eV) & $\omega\gamma$(MeV) \\
\hline

1.685 & 1.104 & 1/2$^+$ & 2 &0.0861 & & & 0.000346 & 26.676 &$4.939\times10^{-11}$ \\
2.178 & 1.597 & 9/2$^+$ & 2 &0.0028 & & 0.3667 & 0.001578 & 1375.825 & $1.127\times10^{-9}$ \\
2.358 & 1.777 & 7/2$^+$ & 0 &0.0269 & 0.0002 & 0.217 &0.002893 & 36270.718 & $1.653\times10^{-9}$\\

\hline
1.685 & 0.940 & 1/2$^+$ & 0 &  &0.1190 & 0.0137& 0.000346 & 606.16  &$1.152\times10^{-10}$ \\
2.178 & 1.430 & 9/2$^+$ &  & & &  & 0.001578 &  &  \\
2.358 & 1.613 & 7/2$^+$ & 2 & 0.0118 &  &  &0.002893 & 45.294 & $3.858\times10^{-9}$\\

\hline \hline
\end{tabular}
\end{table}

For the non-resonant contribution, it is directly related to the effective astrophysical S factor ($S_{\rm eff}$) in the energy range of the Gamow window via the following expression \citep{Sun19}
\begin{equation}
N_A\left\langle\sigma\nu\right\rangle_{dc}=7.8327\times10^9\times\left(\frac{Z_pZ_T}{\mu T_9^2}\right)^{1/3}\times S_{\rm eff}\times {\rm exp}\left[-4.2487\left(\frac{Z_p^2Z_T^2\mu}{T_9}\right)^{1/3}\right]({\rm cm^3s^{-1}mol^{-1}})\text{,}
\end{equation}
where $S_{\rm eff}$ can be parameterized by the formula 
\begin{equation}
S_{\rm eff}\approx S(0)\left[1+0.09807\left(\frac{T_9}{Z_p^2Z_T^2\mu}\right)^{1/3}\right]\text{,}
\end{equation}
and $S(0)$ is the astrophysical S factor at zero energy in units of MeV$\cdot$b.

The $S(0)$ for direct capture into the ground state of $^{27}$S has been calculated with a RADCAP code \citep{Ber03} by using a Woods-Saxon nuclear potential (central + spin-orbit) and a Coulomb potential of a uniform charge distribution. The nuclear potential parameters were determined by matching the bound-state energy. The spectroscopic factors used for the direct capture calculation are taken from the shell model calculation, and the obtained $S(0)$ are 33.67 keV$\cdot$b and 0.68 keV$\cdot$b which corresponds to the direct capture from the ground state and first excited state of $^{26}$P, respectively. Our uncertainty of $S(0)$ for direct capture is set to be 41$\%$, which includes not only an assumed uncertainty of 40$\%$ as in \citep{Downen22} but also the contribution due to the uncertainty of the Q-value. 

In principle, the total reaction rate is the sum of the capture rate on all thermally excited states in the target nucleus weighted with their individual population factors \citep{Schatz05}:
\begin{equation}
N_A\left\langle\sigma\nu\right\rangle = \sum_{i} (N_A\left\langle\sigma\nu\right\rangle_{res\:i} + N_A\left\langle\sigma\nu\right\rangle_{dc\:i}) \frac{(2J_i+1)e^{-E_i/kT}}{\sum_{n}(2J_n+1)e^{-E_n/kT}}
\end{equation}
In this work, we consider only capture on the ground state and the first excited state in $^{26}$P. The $^{26}$P ground state is known to have spin and parity 3$^+$. The spin for the experimentally known first excited state at 164.4 keV has not been determined unambiguously but we assign a spin of 1$^+$ based on the level structure of the $^{26}$Na mirror and our shell model calculations.

\begin{figure}[ht!]
\begin{center}
\includegraphics[width=0.7\textwidth]{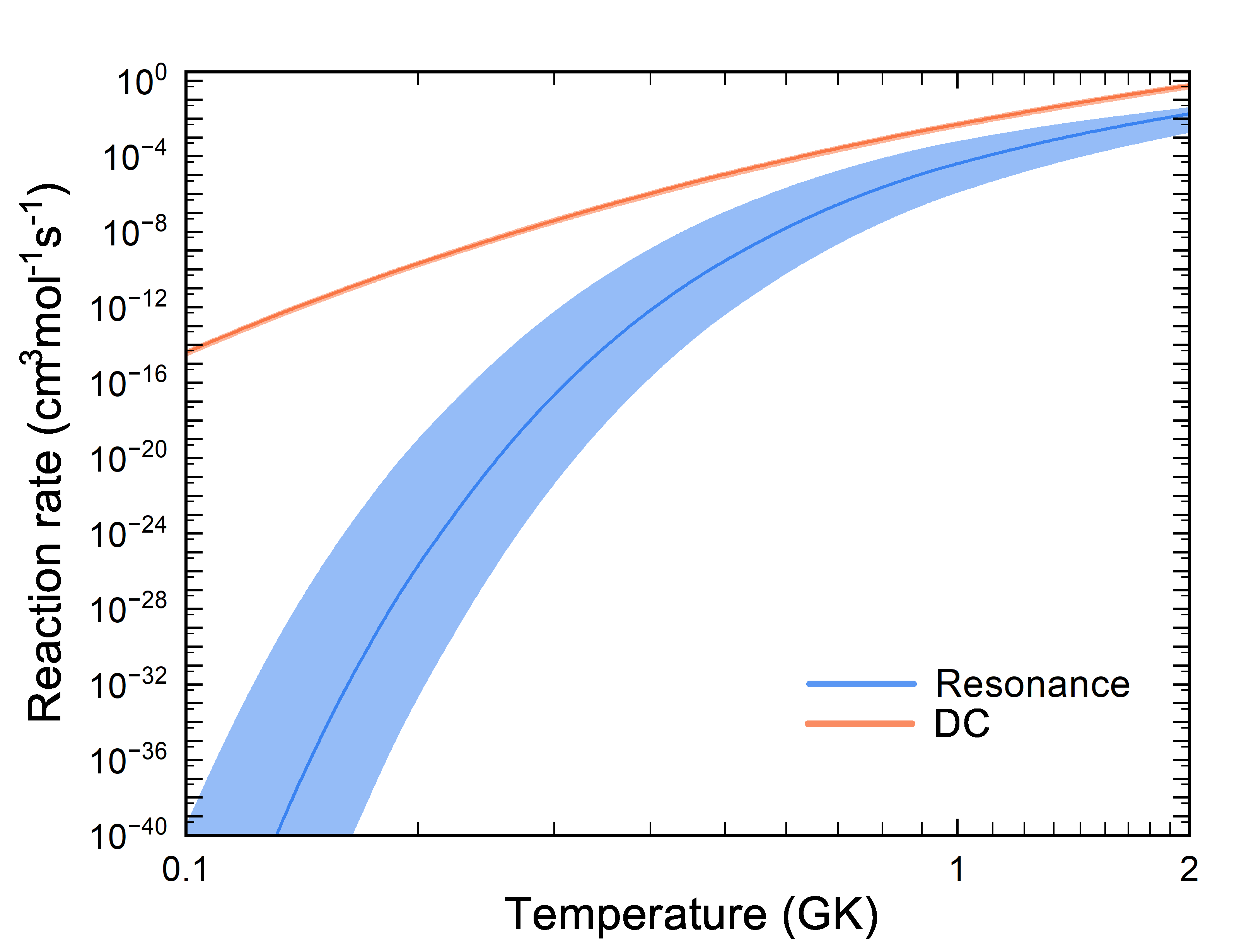}
\end{center}
\caption{\label{fig2} (Color online) Direct capture (DC) and the resonant capture (the sum of the three considered resonances, see the text)} contributions for the thermonuclear $^{26}$P($p$,$\gamma$)$^{27}$S reaction rate  (in units of cm$^{3}$ mol$^{-1}$ s$^{-1}$), and the blue (red) shaded band represents the uncertainties of resonant (direct capture) reaction rates.
\label{reaction_rate}
\end{figure}

We calculate the direct contribution and the total contributions of the three resonances in which the thermalization effect on reaction rate is considered, as shown in Fig.\ref{reaction_rate}. The solid red line is for the direct component with the narrow shallow red band for its 41\% uncertainty. The solid blue line is for the total contribution of the three resonances, and the blue shaded band is the corresponding uncertainty of resonant rate from the resonance energy uncertainty of 271 keV, which is produced in quadrature the uncertainty of $S_p$ and that of the resonance state energy in $^{27}$S. Here, the uncertainty of $^{27}$S resonance state energy is assumed to be 166 keV, which is the maximum energy difference between the shell-model prediction and the experimental measurements for an arbitrary state below 2.5 MeV in $^{27}$Na. We can see clearly that the direct capture makes the most important contribution to the total $^{26}$P($p$,$\gamma$)$^{27}$S reaction rate, while the contributions from resonances are negligible because the resonance energies of the three resonances are all larger than 1 MeV.  The weighted reaction rate is obtained by summing the direct contribution and the resonance contributions after the thermalization correction, as shown in Table \ref{Total_reaction_rate}. The new rate in units of cm$^3$ mol$^{-1}$ s$^{-1}$ can be well fitted (less than 0.33\% error in 0.01-10 GK) by the following analytic expression in the standard seven parameter format of REACLIB:

\begin{equation}
N_A\left\langle\sigma\nu\right\rangle=\mathrm{exp}(21.4303-0.00292768T_9^{-1}-25.0636T_9^{-1/3}-1.69602T_9^{1/3}+0.0866331T_9-0.00201917T_9^{5/3}-0.106906\ln{T_9}).
\end{equation}

\begin{table*}
\scriptsize
\caption{Direct, resonant, and total reaction rates for $^{26}$P($p$,$\gamma$)$^{27}$S based on the present work (in units of cm$^3$ mol$^{-1}$ s$^{-1}$). \label{Total_reaction_rate}}
\begin{tabular}{cccccccc}
\hline \hline
$T_9$(GK) &  \multicolumn{2}{c}{Ground state} & & \multicolumn{2}{c}{First excited state} & & Weighted reaction rate\\
\cline{2-3}  \cline{5-6} 
  & $N_A\left\langle\sigma\nu\right\rangle_{\text{nr}}$ & $N_A\left\langle\sigma\nu\right\rangle_{\text{res}}$  & & $N_A\left\langle\sigma\nu\right\rangle_{\text{nr}}$ & $N_A\left\langle\sigma\nu\right\rangle_{\text{res}}$& & $N_A\left\langle\sigma\nu\right\rangle_{\text{total}}$ \\
\hline

0.1 & 4.09$\times10^{-15}$ & 5.81$\times10^{-54}$ & & 8.25$\times10^{-17}$ & 2.50$\times10^{-45}$ &  & 4.09$\times10^{-15}$  \\
0.2 & 2.18$\times10^{-10}$ & 1.36$\times10^{-26}$ & & 4.39$\times10^{-12}$ & 4.30$\times10^{-22}$ &  & 2.18$\times10^{-10}$  \\
0.3 & 4.14$\times10^{-08}$ & 1.39$\times10^{-17}$ & & 8.34$\times10^{-10}$ & 1.84$\times10^{-14}$ &  & 4.14$\times10^{-08}$  \\
0.4 & 1.12$\times10^{-06}$ & 3.91$\times10^{-13}$ & & 2.25$\times10^{-08}$ & 1.06$\times10^{-10}$ &  & 1.12$\times10^{-06}$  \\
0.5 & 1.15$\times10^{-05}$ & 1.69$\times10^{-10}$ & & 2.32$\times10^{-07}$ & 1.78$\times10^{-08}$ &  & 1.14$\times10^{-05}$  \\
0.6 & 6.80$\times10^{-05}$ & 9.24$\times10^{-09}$ & & 1.37$\times10^{-06}$ & 5.13$\times10^{-07}$ &  & 6.68$\times10^{-05}$  \\
0.7 & 2.80$\times10^{-04}$ & 1.56$\times10^{-07}$ & & 5.63$\times10^{-06}$ & 5.47$\times10^{-06}$ &  & 2.73$\times10^{-04}$  \\
0.8 & 8.94$\times10^{-04}$ & 1.27$\times10^{-06}$ & & 1.80$\times10^{-05}$ & 3.15$\times10^{-05}$ &  & 8.63$\times10^{-04}$  \\
0.9 & 2.38$\times10^{-03}$ & 6.47$\times10^{-06}$ & & 4.80$\times10^{-05}$ & 1.20$\times10^{-04}$ &  & 2.28$\times10^{-03}$  \\
1.0 & 5.54$\times10^{-03}$ & 2.38$\times10^{-05}$ & & 1.12$\times10^{-04}$ & 3.48$\times10^{-04}$ &  & 5.26$\times10^{-03}$  \\
1.5 & 1.07$\times10^{-01}$ & 1.44$\times10^{-03}$ & & 2.15$\times10^{-03}$ & 8.40$\times10^{-03}$ &  & 9.79$\times10^{-02}$  \\
2.0 & 6.77$\times10^{-01}$ & 1.40$\times10^{-02}$ & & 1.36$\times10^{-02}$ & 4.75$\times10^{-02}$ &  & 6.02$\times10^{-01}$  \\
2.5 & 2.50$\times10^{+00}$ & 5.79$\times10^{-02}$ & & 5.03$\times10^{-02}$ & 1.49$\times10^{-01}$ &  & 2.16$\times10^{+00}$  \\
3.0 & 6.71$\times10^{+00}$ & 1.49$\times10^{-01}$ & & 1.35$\times10^{-01}$ & 3.31$\times10^{-01}$ &  & 5.67$\times10^{+00}$  \\
3.5 & 1.47$\times10^{+01}$ & 2.86$\times10^{-01}$ & & 2.96$\times10^{-01}$ & 5.84$\times10^{-01}$ &  & 1.23$\times10^{+01}$  \\
4.0 & 2.80$\times10^{+01}$ & 4.58$\times10^{-01}$ & & 5.64$\times10^{-01}$ & 8.83$\times10^{-01}$ &  & 2.28$\times10^{+01}$  \\
5.0 & 7.65$\times10^{+01}$ & 8.48$\times10^{-01}$ & & 1.54$\times10^{+00}$ & 1.52$\times10^{+00}$ &  & 6.05$\times10^{+01}$  \\

\hline \hline
\end{tabular}
\end{table*}

\begin{figure}[ht!]
\begin{center}
\includegraphics[width=0.6\textwidth]{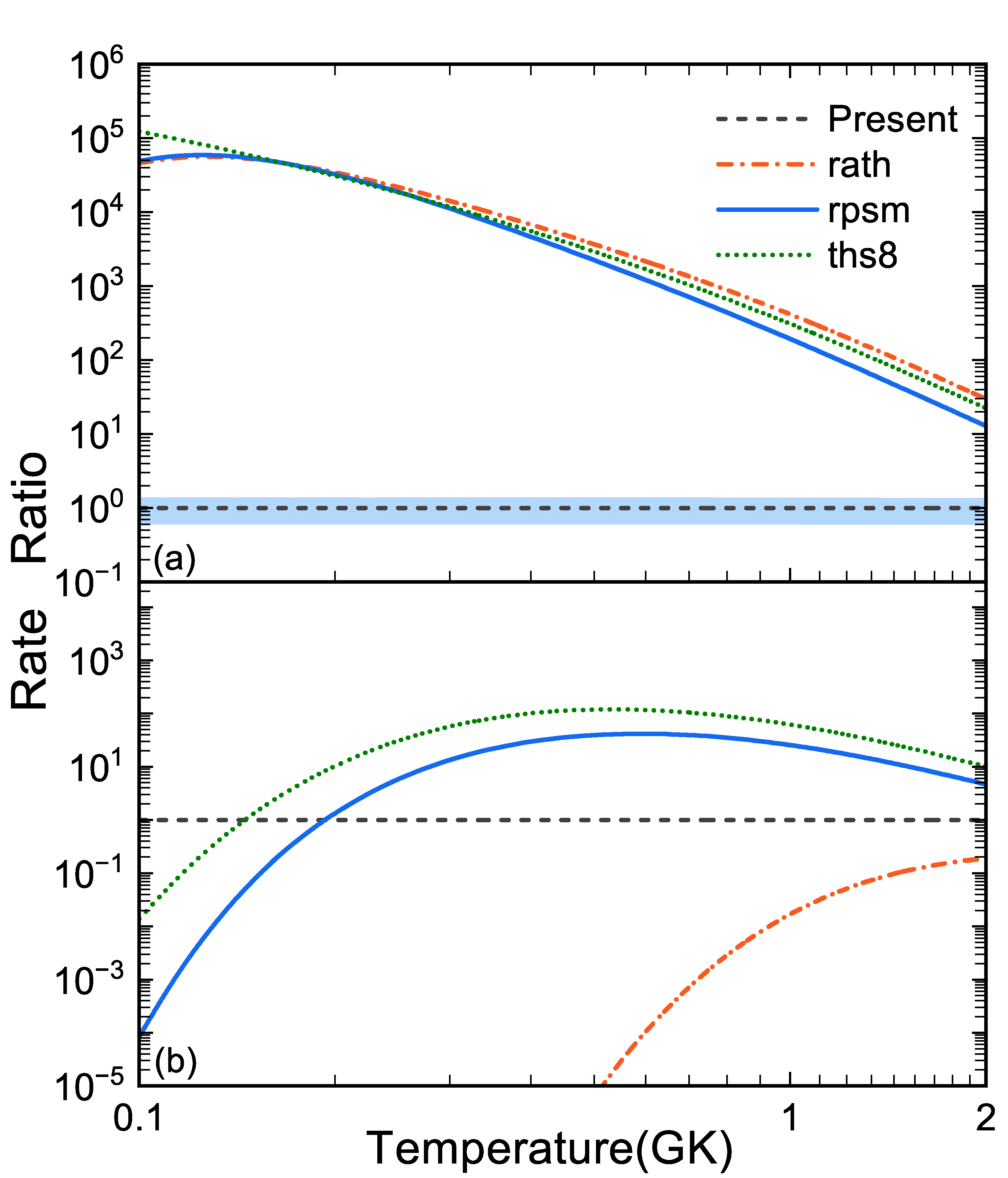}
\end{center}
\caption{\label{fig4} (Color online) Ratios of the $^{26}$P($p$,$\gamma$)$^{27}$S rate from different sources (rath \citep{Rauscher00}, rpsm \citep{Rauscher1999}, ths8 \citep{Cyburt10}) to the present rate. (a) Ratios for the forward reaction rate; (b) Ratios for the reverse reaction rate.
\label{forward_reverse_ratio}}
\end{figure}

For $^{26}$P($p$,$\gamma$)$^{27}$S reaction, we know its reverse reaction rate (also called photodisintegration rate) can be calculated directly by the expression \citep{Iliadis15}

\begin{equation}
\label{Rev_rate}
\lambda_{(\gamma,p)}=\frac{2G_f}{G_i}\left(\frac{\mu kT}{2\pi\hbar^2}\right)^{3/2}\mathrm{exp}\left(-\frac{Q_{(p,\gamma)}}{kT}\right)\left\langle\sigma v\right\rangle
\end{equation}
where $G_i$ and $G_f$ are the partition functions of the initial and final nuclei. As mentioned above, the ratio of the ($\gamma,p$) reaction rate to the ($p,\gamma$) rate depends exponentially on $Q_{(p,\gamma)}$ and is therefore very sensitive to nuclear masses. Using the new Q-value determined by the new $^{27}$S mass, we recalculate the $^{27}$S($\gamma$,$p$)$^{26}$P rate. 
In Fig.\ref{forward_reverse_ratio} we plot the new reaction rate of $^{26}$P($p$,$\gamma$)$^{27}$S and the uncertainties. The other versions (rath, rpsm, ths8) collected in JINA REACLIB are also added for comparison. Forward reaction rates are shown in Panel(a), while reverse rates are shown in Panel(b). It can be seen that the newly obtained forward rate is smaller than all of the previous three rates. For reverse process $^{27}$S($\gamma$,$p$)$^{26}$P, the new rate is also smaller than those from rpsm and ths8 over the temperature range of 0.2-2 GK but is larger than the rath rate, which is due to the rather large Q-value of 1.452 MeV.

\section{Astrophysical implications} \label{sec:Astrophysical}
\begin{figure}[ht!]
\begin{center}
\includegraphics[width=0.7\textwidth]{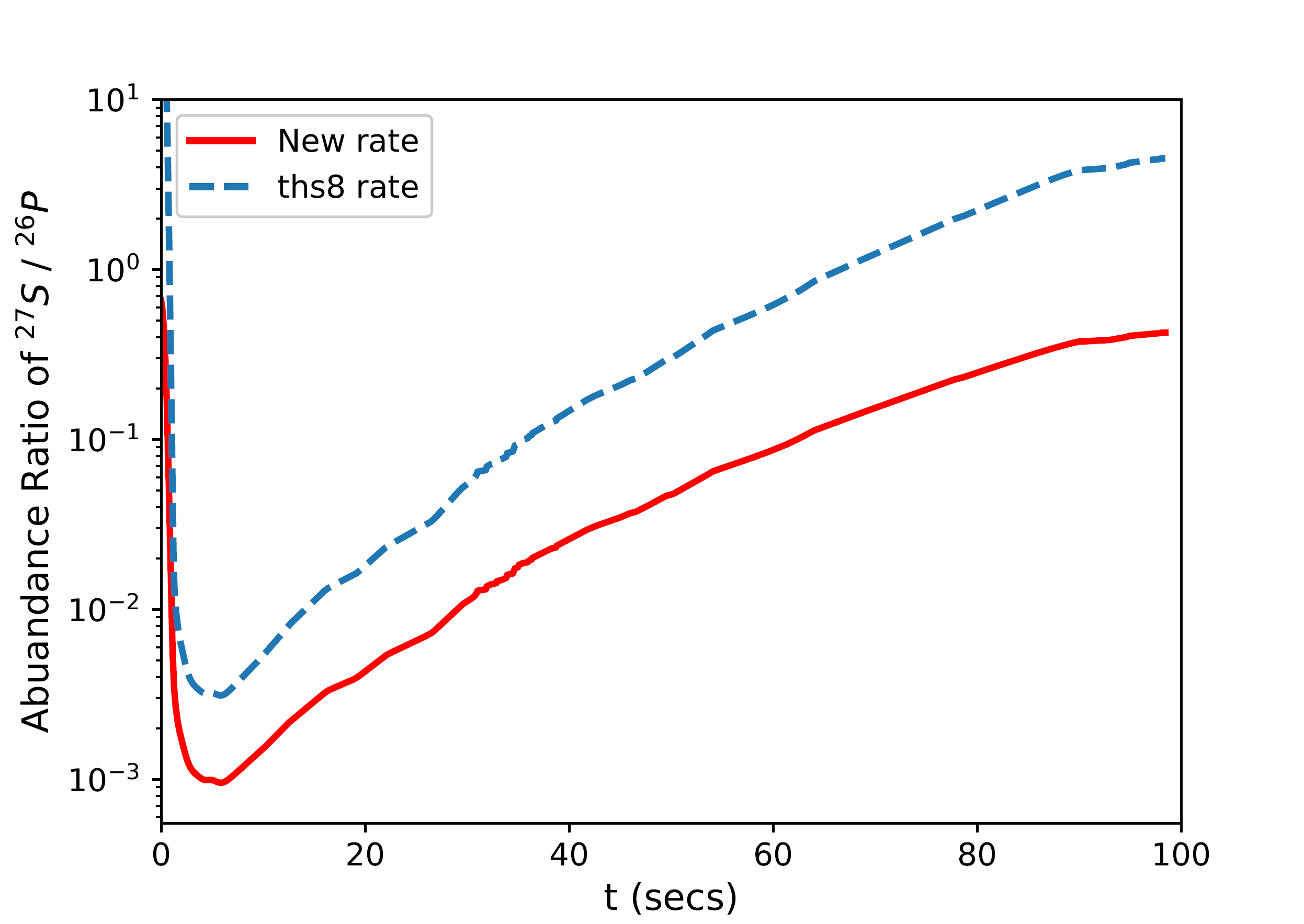}
\end{center}
\caption{(Color online) The abundance ratio of $^{27}$S/$^{26}$P compared using the newly determined rate in this work and for the ths8 rate, plotted as a function of time.
\label{abu_S27_to_P26_ratio}}
\end{figure}

\begin{figure}[htbp]
    \begin{minipage}[t]{0.5\linewidth}
        \centering
        \includegraphics[width=9.2cm]{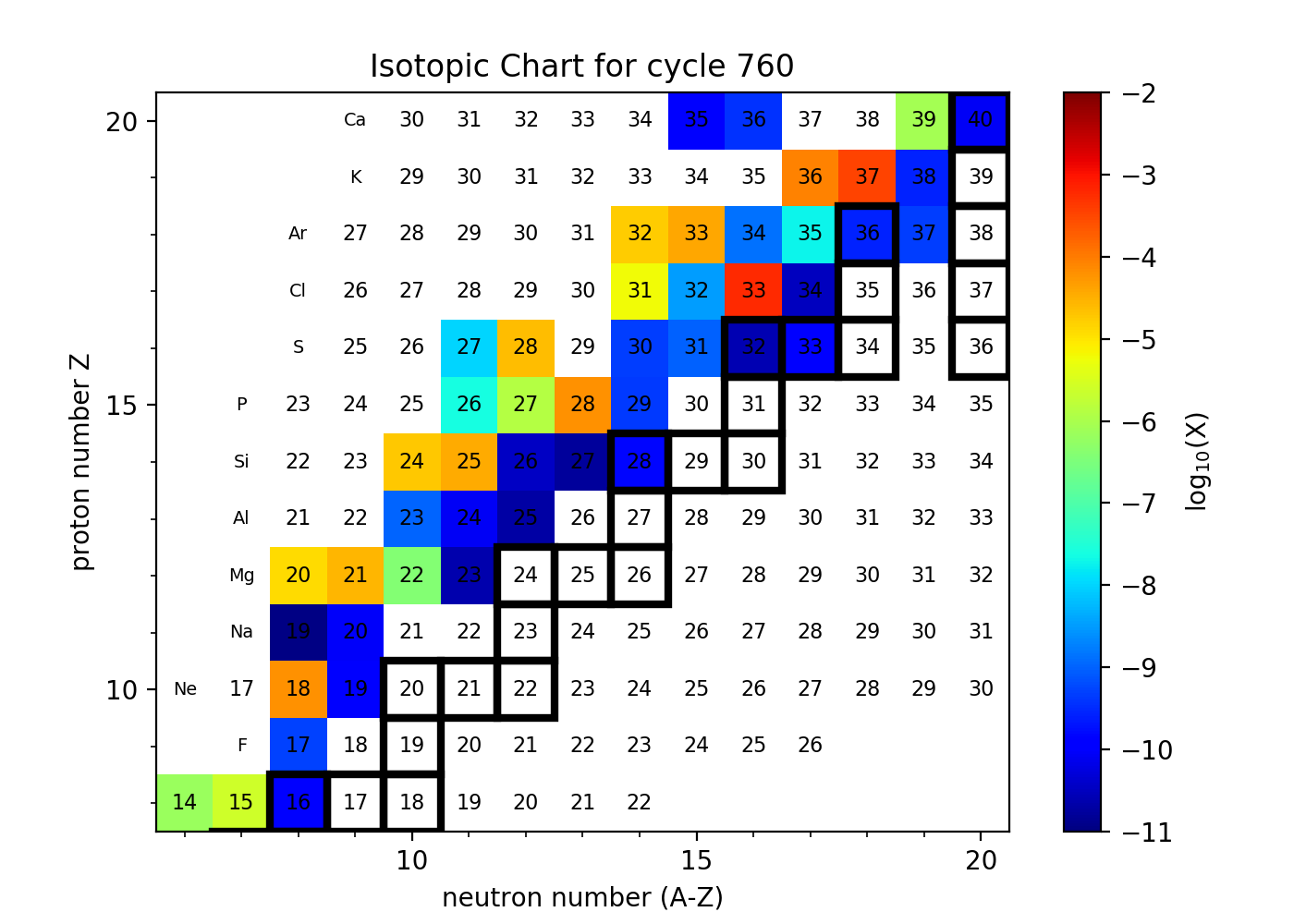}
        \centerline{(a)}
    \end{minipage}%
    \begin{minipage}[t]{0.5\linewidth}
        \centering
        \includegraphics[width=9.2cm]{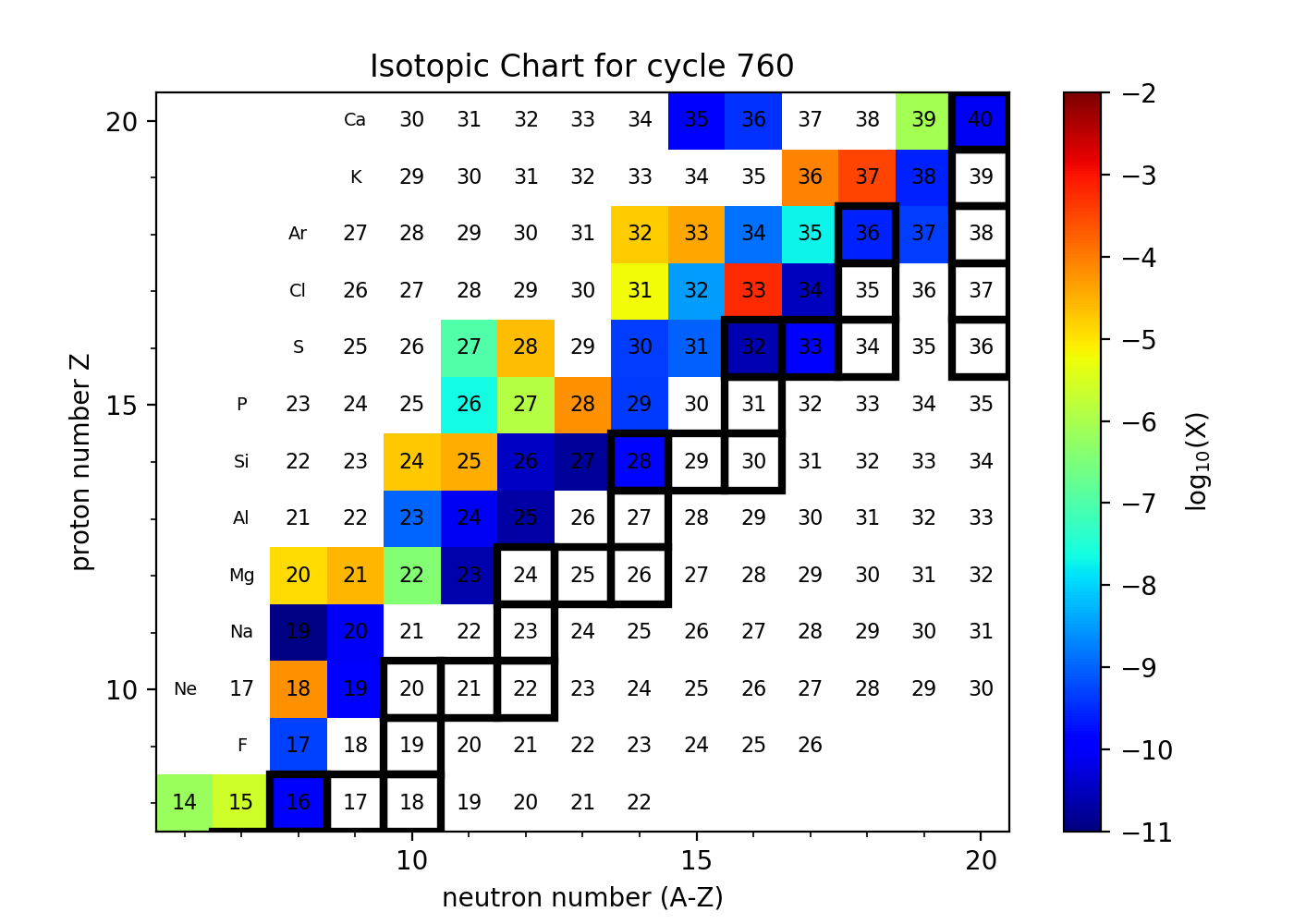}
        \centerline{(b)z}
    \end{minipage}
    \caption{(Color online) The distribution of isotope abundances at cycle 760, which corresponds to the termination time of the rp-process. (a) is for the case using the new forward and reverse rates of $^{26}$P($p$,$\gamma$)$^{27}$S, while (b) is for the case adopting ths8 rates.}
    \label{abu_chart_760}
\end{figure}

 In order to explore the impact of the new forward and reverse rates for $^{26}$P($p$,$\gamma$)$^{27}$S on the rp-process, we make use of the one-zone post-processing nucleosynthesis code ppn, a branch of the NuGrid framework \citep{NuGrid08,Pavel14}. Both the initial composition of the accreted material($X\rm^0_H$= 0.735, $X\rm^0_{^4He}$= 0.245, $X\rm^0_{^{14}O}$ = 0.007 and $X\rm^0_{^{15}O}$ = 0.013) and the profile used in the simulation are taken from \cite{Koike04}. We perform two simulations with identical nuclear physics and model inputs, but different forward and reverse reaction rates for $^{26}$P($p$,$\gamma$)$^{27}$S. In the first run, we use the $^{26}$P($p$,$\gamma$)$^{27}$S rates from Cyburt et al. \citep{Cyburt10} (labeled as ths8 in JINA REACLIB), while the newly obtained forward and reverse rates for this reaction are used in the second run. The calculated results show that the new rates can sensitively affect the abundance ratio of $^{27}$S/$^{26}$P in X-ray burst. It can be seen clearly in Fig.\ref{abu_S27_to_P26_ratio} that the obtained $^{27}$S/$^{26}$P ratio using the new rates is overall smaller than that adopting the ths8 rates by about a factor of 10. The reason for this is twofold. On one side, this is because the new $^{26}$P($p$,$\gamma$)$^{27}$S rate is much smaller than the ths8 rate, resulting in less direct production of $^{27}$S. On the other side, the new $S_p$ of 581 keV is also smaller than that of 719 keV from ths8, implying a relatively stronger level of photodisintegration which prevents the synthesis of $^{27}$S. Fig.\ref{abu_chart_760} shows the final isotope abundance distributions for the case using the new rates(panel (a)) and that adopting ths8 rates(panel (b)), respectively. It is clear that the accumulated $^{27}$S abundance in Fig.\ref{abu_chart_760} (a) is smaller than that in Fig.\ref{abu_chart_760} (b). In other words, for the case of adoption of our new rates, the reaction flow through the branch of $^{26}$P($p$,$\gamma$)$^{27}$S($\beta^+$,$\nu$)$^{27}$P is much smaller compared with the case of using old rates. In order to investigate whether the uncertainty of 214 keV of the new $S_p$ will change this conclusion, we also perform additional two runs in which only the photodisintegration rates are changed by taking into account the uncertainty of the new $S_p$ for $^{27}$S, while all other rates remain the same as those used in the second run.  It is known that the uncertainty of the new $S_p$ is determined by the uncertainties from the nuclear masses of $^{26}$P and $^{27}$S. Thus, the first run is for the case considering ${S_p}$ upper limit (marked as $\overline{S_p}$), which corresponds to the case of adopting the upper limit of the $^{26}$P mass and the lower limit of the $^{27}$S mass in the process of calculating the respective photodisintegration rate of $^{26}$P($\gamma$,$p$) and $^{27}$S($\gamma$,$p$). Similarly, the second run is for the case of using the ${S_p}$ lower limit (marked as $\underline{S_p}$), in which the lower limit of the $^{26}$P mass and the upper limit of the $^{27}$S mass are used to derive the respective reverse rates. Our calculations show the final accumulated material on $^{27}$S through $^{26}$P($p$,$\gamma$)$^{27}$S for both cases of $\overline{S_p}$ and $\underline{S_p}$ decreases relative to the case of using $S_p$ -- by about 25\% and 39\%, respectively. The production of $^{27}$S is reduced when using the former due to the enhanced $^{26}$P($\gamma$,$p$)$^{25}$Si rate from assuming the upper limit of the $^{26}$P mass, which effectively inhibits the reaction flow passing through $^{26}$P.  In the case of the latter, using the lower limit of ${S_p}$ results in a stronger photodisintegration rate of $^{27}$S($\gamma$,$p$)$^{26}$P, directly preventing the reaction flow to $^{27}$S. 

\begin{figure}[ht!]
\begin{center}
\includegraphics[width=0.7\textwidth]{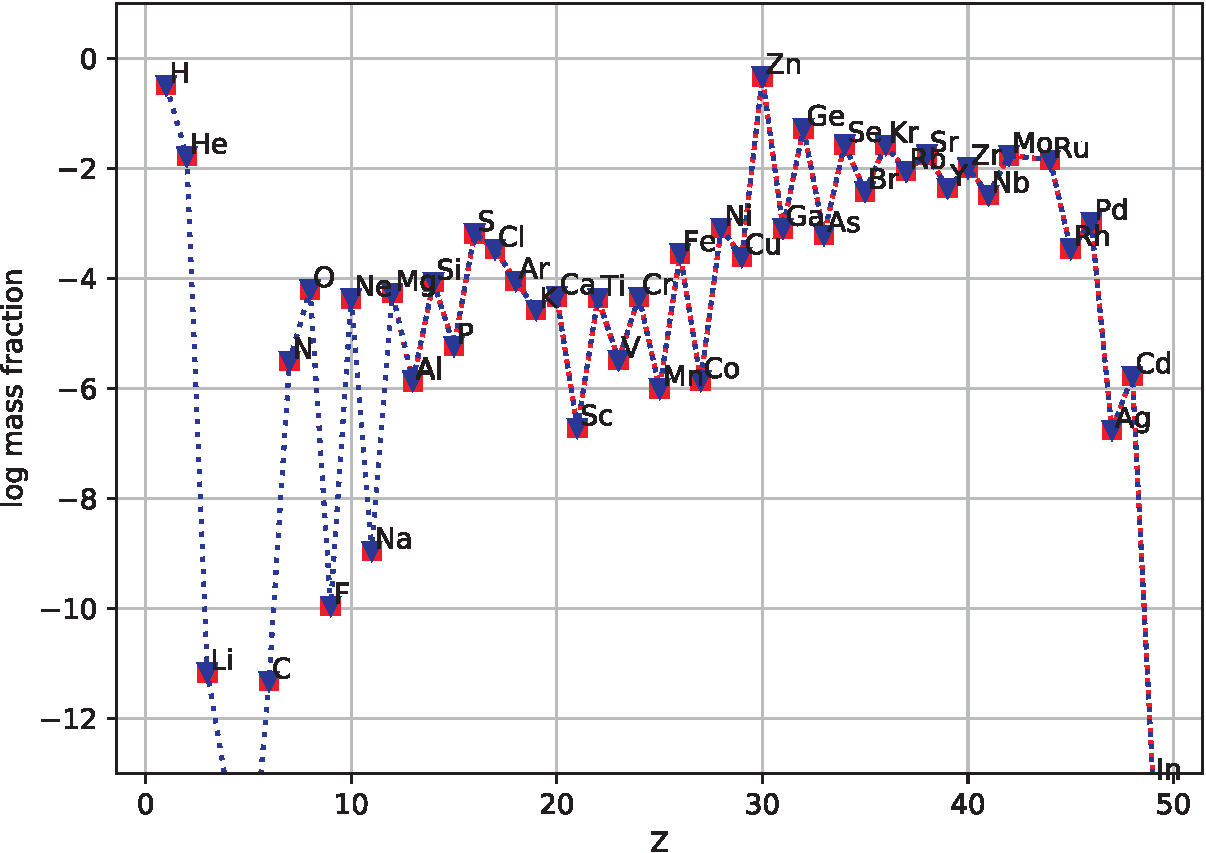}
\end{center}
\caption{(Color online) The final elemental abundance after full decay of all the unstable isotopes. The solid red square is for the case using the new rates for $^{26}$P($p$,$\gamma$)$^{27}$S, while the blue triangle is for that adopting the ths8 rates.
\label{abu_decay}}
\end{figure}

In light of the variation of the accumulated material on $^{27}$S when assuming different forward and reverse rates of $^{26}$P($p$,$\gamma$)$^{27}$S, we compare the decayed elemental abundances(i.e., accounting for the complete contribution from the radioactive decay of all unstable isotopes) calculated using the new rates with those calculated using the ths8 rates. The results are plotted in Fig.\ref{abu_decay} in which the solid red squares are the abundances calculated using our new rates and the blue triangle corresponds to that of adopting ths8 rates. It can be seen that using the new rates did not result in  a significant change in elemental abundance. The largest difference is seen for aluminum ($^{27}$Al), with a decrease of 7.1\%.

\section{Conclusion} \label{sec:conclusion}
In this study, we employ the shell-model to predict the energy structure information of $^{27}$S. In addition, using the updated proton threshold determined by the new $^{27}$S mass, we recalculate the thermonuclear reaction rates for the $^{26}$P($p$,$\gamma$)$^{27}$S reaction, which is important in the rp-process. The calculated result shows the direct contribution dominates this reaction rate in comparison to the resonance contribution. Both our new rates for forward and reverse reactions are different from the other three rates from JINA REACLIB. The ratio of the old rates to our new rate can even be up to a factor of 5 orders of magnitude at the temperature of 0.1 GK. At the stellar conditions relevant to this study (0.4 GK\textless $T_9$ \textless1.35 GK), the typical variation of reaction rate is in the order of 2 to 4. We investigate the effect of the new thermonuclear reaction rates on the nucleosynthesis in rp-process using the ppn post-processing code. It is found that the ratio of isotope abundances of $^{27}$S/$^{26}$P when adopting the new rates is smaller by a factor of 10 than that using the ths8 rates from the JINA database. In addition, the accumulated material on the $^{26}$P nucleus is larger than that on $^{27}$S during the whole rp-process episode, which is markedly different from the result of using the old rates of $^{26}$P($p$,$\gamma$)$^{27}$S. For the flow reaching branch point nucleus $^{26}$P, the reaction chain of $^{26}$P($p$,$\gamma$)$^{27}$S($\beta^+$,$\nu$)$^{27}$P competes with the $\beta^+$ branch of $^{26}$P($\beta^+$,$\nu$)$^{26}$Si($p$,$\gamma$)$^{27}$P. Our calculations confirm that $^{26}$P($p$,$\gamma$)$^{27}$S($\beta^+$,$\nu$)$^{27}$P is not the major path for the synthesis of $^{27}$P after adopting our new forward and reverse rates.  The adoption of the new reaction rates for $^{26}$P($p$,$\gamma$)$^{27}$S only reduces the final production of aluminum by 7.1\%, and has no discernible impact on the yield of other elements.

\begin{acknowledgments}

This work was financially supported by the National Key R\&D Program of China Grant no. 2022YFA1603300, the Strategic Priority Research Program of Chinese Academy of Sciences Grant No. XDB34020204, and the Youth Innovation Promotion Association of Chinese Academy of Sciences under Grant No. 2019406, and in part by the National Science Foundation under Grant No. OISE-1927130 (IReNA); the National Natural Science Foundation of China under Grant No. 12205340;  the Gansu Natural Science Foundation under Grant No. 22JR5RA123. This research was made possible by using the computing resources of Gansu Advanced Computing Center. C.A.B. acknowledges support by the U.S. DOE Grant DE-FG02-08ER41533. MP and TT acknowledge support from STFC (through the University of Hull’s Consolidated Grant ST/R000840/1), and access to viper, the
University of Hull HPC Facility. We acknowledge the support
to NuGrid from the National Science Foundation (NSF, USA) under
grant No. PHY-1430152 (JINA Center for the Evolution of the Elements). MP thanks the "Lendulet-2014" Program of the Hungarian Academy
of Sciences (Hungary), the ERC Consolidator Grant funding scheme
(Project RADIOSTAR, G.A. n. 724560, Hungary), the ChETEC
COST Action (CA16117), supported by the European Cooperation
in Science and Technology, and the IReNA network supported by
NSF AccelNet. MP acknowledges support from the ChETEC-INFRA
project funded by the European Union’s Horizon 2020 Research and
Innovation programme (Grant Agreement No 101008324). 

\end{acknowledgments}

%




\bibliography{reference}{}
\bibliographystyle{aasjournal}

\end{document}